\documentclass[prl,twocolumn,superscriptaddress,showpacs]{revtex4}
\usepackage{mathrsfs}
\usepackage{amssymb}
\usepackage{amsmath}
\usepackage{graphicx}

\usepackage{bm}

\begin{document}

\title{Direct experimental verification of quantum commutation relations for Pauli operators}

\author{Xing-Can Yao}
\affiliation{Hefei National Laboratory for Physical Sciences at
Microscale and Department of Modern Physics, University of Science
and Technology of China, Hefei, Anhui 230026, China}

\author{Jarom\'{\i}r Fiur\'{a}\v{s}ek}
\affiliation{Department of Optics, Palack\'{y} University, 17.
listopadu 12, 77146 Olomouc, Czech Republic}

\author{He Lu} \affiliation{Hefei National Laboratory for Physical
Sciences at Microscale and Department of Modern Physics, University
of Science and Technology of China, Hefei, Anhui 230026, China}

\author{Wei-Bo Gao}
\affiliation{Hefei National Laboratory for Physical Sciences at
Microscale and Department of Modern Physics, University of Science
and Technology of China, Hefei, Anhui 230026, China}

\author{Yu-Ao Chen} \affiliation{Hefei National Laboratory for
Physical Sciences at Microscale and Department of Modern Physics,
University of Science and Technology of China, Hefei, Anhui 230026,
China}

\author{Zeng-Bing Chen}
\affiliation{Hefei National Laboratory for Physical Sciences at
Microscale and Department of Modern Physics, University of Science
and Technology of China, Hefei, Anhui 230026, China}

\author{Jian-Wei Pan}
\affiliation{Hefei National Laboratory for Physical Sciences at
Microscale and Department of Modern Physics, University of Science
and Technology of China, Hefei, Anhui 230026, China}
\affiliation{Physikalisches Institut, Ruprecht-Karls-Universit\"{a}t
Heidelberg, Philosophenweg 12, 69120 Heidelberg, Germany}

\begin{abstract}
We propose and demonstrate scheme for direct experimental testing of
quantum commutation relations for Pauli operators. The implemented
device is an advanced quantum processor that involves two
programmable quantum gates. Depending on a state of two-qubit
program register, we can test either commutation or anti-commutation
relations. 
Very good agreement between theory and experiment is observed,
indicating high-quality performance of the implemented quantum
processor and reliable verification of commutation relations for
Pauli operators.
\end{abstract}


\maketitle

 Quantum theory associates each observable physical quantity with Hermitian
operator acting on Hilbert space of states of a given physical
system \cite{Peres95}. A fundamental property of operators
representing different quantities such as position, momentum or
angular momentum of a particle is that they do not mutually commute,
which gives rise to peculiar quantum effects like Heisenberg
uncertainty relations. Although predictions of
quantum theory have been corroborated by countless experiments,
direct observation of the non-commutativity of the underlying
operators has eluded us. Recently, testing of commutation rules for
bosonic creation and annihilation operators
 based on  the combination of
single-photon addition \cite{Zavatta04} and subtraction \cite{Ourjoumtsev06,Neergaard-Nielsen06,Wakui07}
has been reported \cite{Parigi07,Kim08,Zavatta09}.

 Besides non-commutativity of bosonic creation and annihilation operators, commutation rules of Pauli
 operators are also fundamental and important. The Pauli operators were originally
introduced to describe Cartesian components of electron spin. More
generally, they form, together with the identity operator $I$, a
complete basis in the space of operators acting on a two-level
quantum system, a qubit in the language of quantum information
theory \cite{Nielsen00}.
 Pauli operators are both Hermitian and unitary, which can be represented by $2\times2$
 matrices,
\[
X=\left(
\begin{array}{cc}
0 & 1 \\
1 & 0 \\
\end{array}
\right), \quad
Y=\left(
\begin{array}{cc}
0 & -i \\
i & 0 \\
\end{array}
\right), \quad
Z=\left(
\begin{array}{cc}
1 & 0 \\
0 & -1 \\
\end{array}
\right).
\]
 They satisfy fundamental commutation relations characteristic of Lie algebra $\mathrm{su}(2)$,
\begin{equation}
[X,Y]=2iZ,\qquad [Y,Z]=2iX,\qquad [Z,X]=2iY,
\label{commutator}
\end{equation}
where $[A,B]=AB-BA$. Any two different Pauli operators anti-commute, which means that
\begin{equation}
\{X,Y\}=\{Y,Z\}=\{Z,X\}=0,
\label{anticommutator}
\end{equation}
 where $\{A,B\}=AB+BA$. The nonzero commutator implies that, e.g., $ZX \neq XZ$, the overall operation
depends on the order of $X$ and $Z$. By combining Eqs.
(\ref{commutator}) and (\ref{anticommutator}) we find that
$ZX=-XZ=iY$. It follows that, in contrast to the case of bosonic
creation and annihilation operators \cite{Parigi07}, it is
impossible to demonstrate the non-commutativity of $Z$ and $X$  by
applying the two different sequences of operations $ZX$ and $XZ$ to
some input single qubit state $|\psi\rangle$. The output states
$ZX|\psi\rangle=iY|\psi\rangle$ and $XZ|\psi\rangle=-iY|\psi\rangle$
differ only by an overall phase that is not directly observable.
Even applying the operation to a part of an entangled two-qubit
state does not help.

In this letter, we report on the direct experimental verification of
commutation relations for Pauli operators. Our optical scheme
combines two programmable quantum gates
\cite{Nielsen97,Vidal02,Micuda08} and an auxiliary  maximally
entangled two-photon Bell state \cite{Kwiat95,Pan08} to directly
implement the (anti)-commutator of two Pauli operators. We have
completely characterized the commutators by quantum process
tomography \cite{Poyatos97,OBrien04,Lobino08}. Our work
directly reveals the peculiar algebraic structure underlying quantum
theory and represents  realization of an advanced quantum
information processor \cite{Nielsen97}.

  We work with optical qubits encoded into
polarization states of single photons whose Hilbert space is spanned
by the basis states $|H\rangle$ and $|V\rangle$, representing
linearly horizontally and vertically polarized photon, respectively.
In Dirac notation we have $X=|H\rangle\langle V|+|V\rangle\langle
H|$, $Y=i|V\rangle\langle H|-i|H\rangle\langle V|$ and
$Z=|H\rangle\langle H|-|V\rangle\langle V|$. As the Pauli operators
are unitary, they can be deterministically implemented by (a
sequence of) optical wave-plates \cite{Reck94}. Our main tool is a
programmable quantum gate \cite{Nielsen97,Vidal02} where operation
on signal qubit is controlled by the state of program qubit. The
employed linear optical gate \cite{Pittman01,Pittman02,Micuda08}
consists of a polarizing beam splitter (PBS), where signal and
program photons interfere, and the projection of the output program
photon onto diagonally linearly polarized state
$|D\rangle=\frac{1}{\sqrt{2}}(|H\rangle+|V\rangle)$. If the program
photon is prepared in state $|D\rangle$, the polarization state of
the signal photon will be unchanged and hence identity operation is
applied. However, if we prepare program photon in orthogonal state
$|A\rangle=\frac{1}{\sqrt{2}}(|H\rangle-|V\rangle)$, the operation
$Z$ will be applied to signal photon. The success probability of the
gate is $\frac{1}{4}$ and does not depend on the input state of
signal photon.

 In our experiment we combine two programmable quantum gates
with intermediate unitary operation $U$ on the polarization state of
signal photon (see inset in Fig. 1). By preparing the two program
photons in the maximally entangled singlet polarization state
$|\Psi^{-}\rangle=\frac{1}{\sqrt{2}}(|A\rangle|D\rangle-|D\rangle|A\rangle)\equiv
\frac{1}{\sqrt{2}}(|H\rangle|V\rangle-|V\rangle|H\rangle)$ we obtain
the following transformation of state of signal photon,
 \[
 \frac{1}{4\sqrt{2}}\,(ZUI-IUZ)= \frac{1}{4\sqrt{2}}\,[Z,U],
 \]
 which is, up to a constant prefactor, equal to the commutator $C_{Z,U}=[Z,U]$. 
 By varying $U$ we can thus directly test various commutation relations.  
 Moreover, by preparing the two-photon program state in the triplet Bell state
 $|\Phi^{-}\rangle=\frac{1}{\sqrt{2}}(|D\rangle|A\rangle+|A\rangle|D\rangle)\equiv\frac{1}{\sqrt{2}}(|H\rangle|H\rangle-|V\rangle|V\rangle) $ 
 we realize anti-commutator of $Z$ and $U$.

  The experimental setup is shown in Fig. 1. We first generate two pairs of entangled photons by spontaneous down
  conversion. The photons pass through the half-wave and quarter-wave plates (HWPs
and QWPs) and are superposed on the PBSs (see Fig. 1) to implement
the desired quantum gates. To achieve good spatial and
temporal overlap, the photons are spectrally filtered ($\Delta
\lambda_{\mathrm{FWHW}}=3.2~\mathrm{nm}$) and detected by the fiber-coupled
single-photon detectors \cite {Pan08}.

We have experimentally characterized the commutator $C_{Z,U}$ by
full quantum process tomography \cite{Poyatos97,OBrien04,Lobino08}
for the following five different $U$: $X$, $Y$,
$\frac{X+Y}{\sqrt{2}}$, $\frac{Y+Z}{\sqrt{2}}$, and $H$, where
$H=\frac{X+Z}{\sqrt{2}}$ denotes a Hadamard operation.  We have
reconstructed the completely positive map $\chi$ that describes the
transformation of density matrix $\rho$,
\[
\rho_{\mathrm{out}}=\chi(\rho_{\mathrm{in}})=C_{Z,U}\rho_{\mathrm{in}}C_{Z,U}^\dagger.
\]
 We can see that  $\chi$ contains complete information on the commutator $C$.
According to Choi-Jamiolkowski isomorphism
\cite{Jamiolkowski72,Choi75}, $\chi$ can be represented by a
positive semidefinite operator on a Hilbert space of two qubits.
This operator possesses an intuitive and appealing physical
interpretation: it is proportional to a density matrix of a
two-qubit state obtained by sending through the channel $\chi$ one
part of a maximally entangled Bell state
$|\Phi^{+}\rangle=\frac{1}{\sqrt{2}}(|HH\rangle+|VV\rangle)$. We
have probed each commutator by six different input states forming
three mutually unbiased bases  $\{|H\rangle,|V\rangle\}$,
$\{|D\rangle,|A\rangle\}$, and $\{|R\rangle,|L\rangle\}$, where
$|R\rangle$ and $|L\rangle$ denote right- and left-hand circularly
polarized states.
 Each output polarization state was fully characterized by  measurements in those three  bases.
  From the collected data the operator $\chi$ was reconstructed by means of a standard maximum likelihood estimation algorithm \cite{Jezek03}.

\begin{figure}[!t!]
 \includegraphics[width=\linewidth]{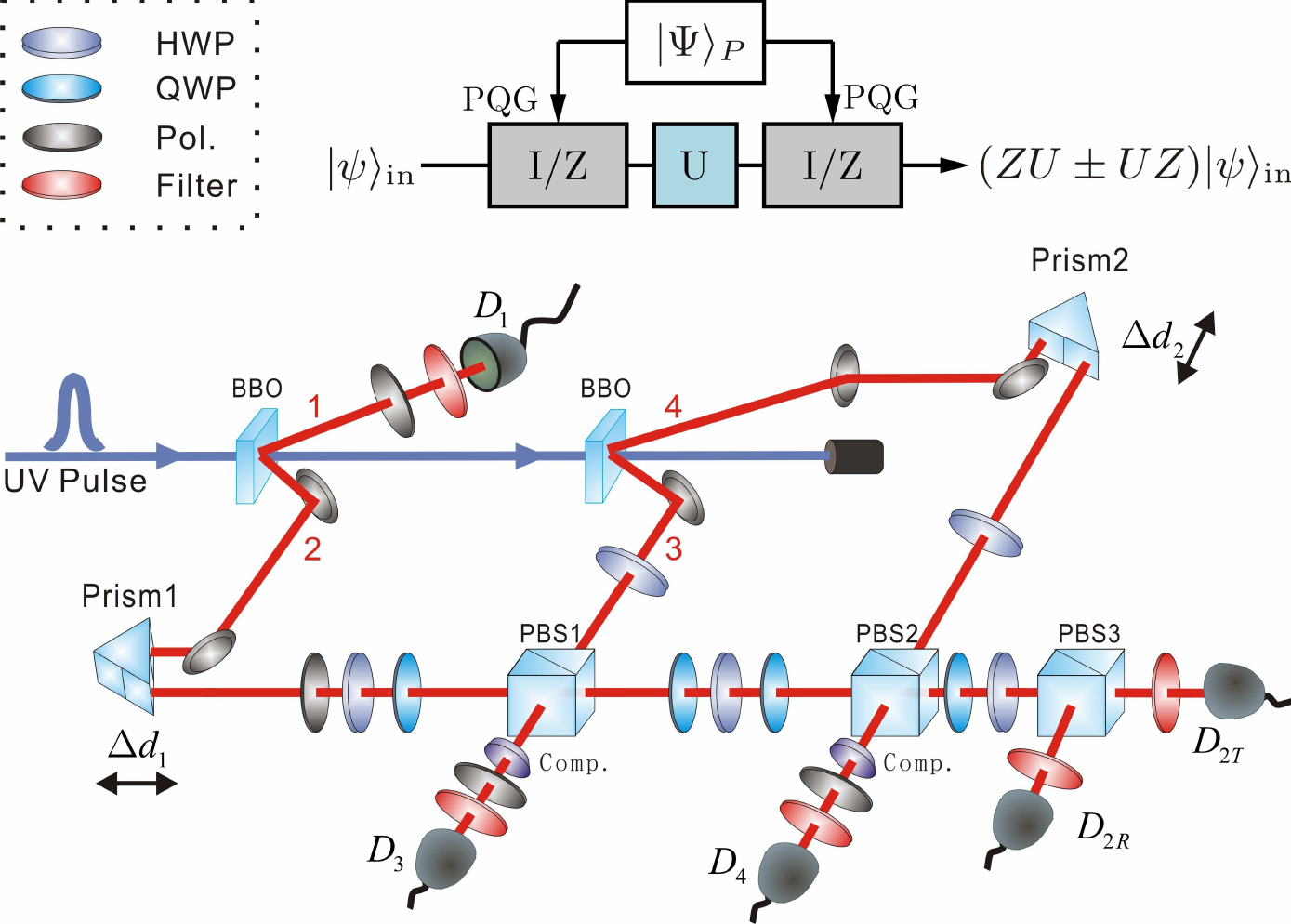}
 \caption{Experimental setup. Femtosecond laser pulses (394nm, 150fs, 76MHz) pass through two main
BBO crystals (2mm) to produce two pairs of entangled photons with an
average count of $2.6 \times 10^4 /s$. The mutual delays between
path 2, 3, 4 are controlled by movable prisms $\Delta d_1$ and
$\Delta d_2$. We incorporate after each PBS a compensator (Comp.) to
counter the additional phase shifts of the PBS. The photon in mode
$2$ serves as signal photon and the photon pair in modes $3$ and $4$
represents the two-qubit program. The inset shows a conceptual
diagram of the core parts of the scheme involving two programmable
quantum gates (PQG) interspersed with another transformation $U$ on
the signal qubit.}
 \end{figure}

\begin{figure}[!t!]
 \includegraphics[width=0.98\linewidth]{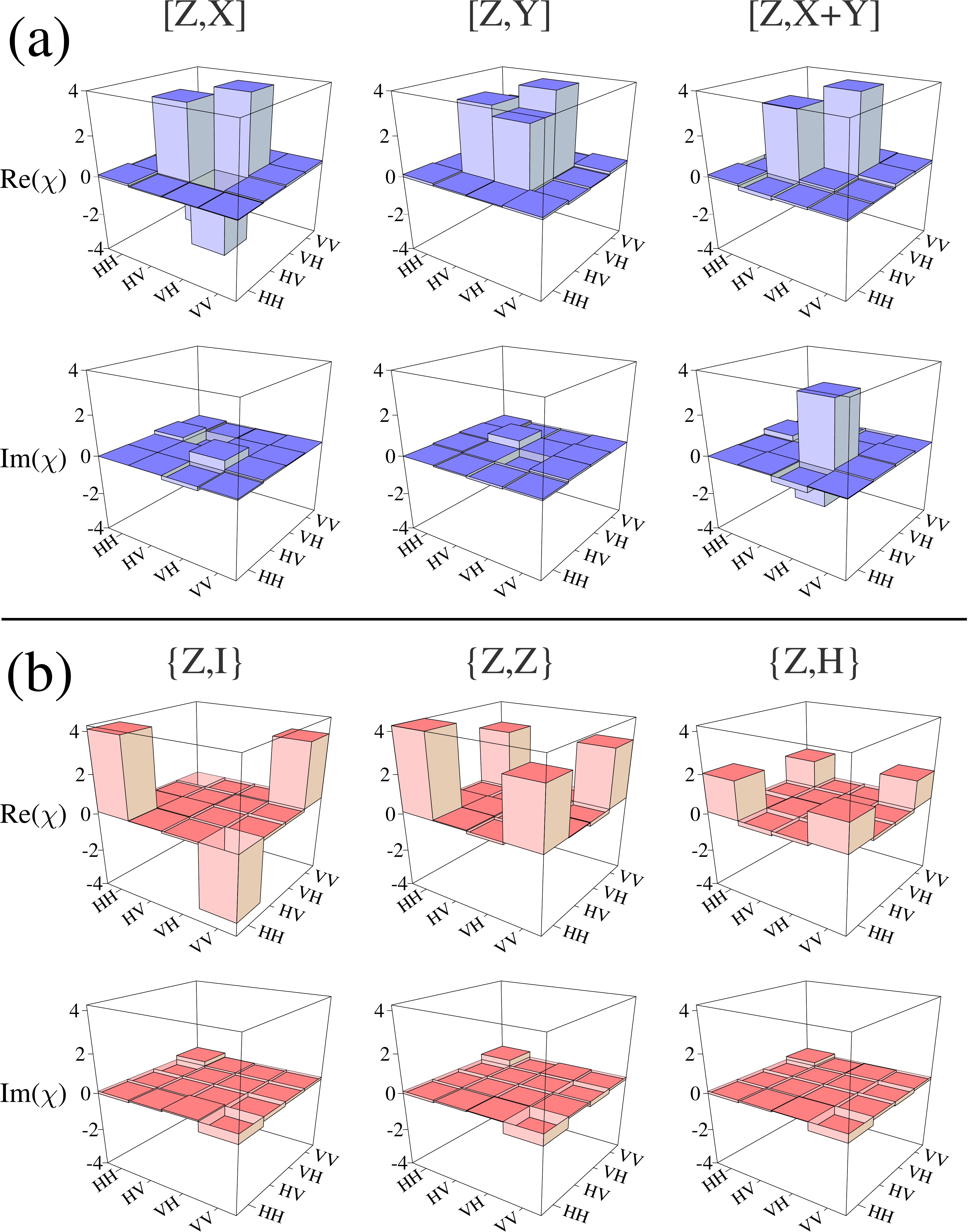}
 \caption{Experimentally determined commutation and anti-commutation operations. Plotted are the reconstructed completely positive maps $\chi$ characterizing the tested commutation relations (a) and anti-commutation relations (b). For each case both real and imaginary part of the $4\times4$ Hermitian matrix $\chi$ is shown. The matrices are normalized according to the experimentally evaluated $K$ factors given in Tables I and II (see main text).}
 \end{figure}

Examples of the results are shown in Fig. 2a where we plot the real
and imaginary parts of the reconstructed $\chi$ for three different
$U$. For all five tested $U$ the theory predicts that $C_{Z,U}=KV$
where $V$ is a unitary operation and $K$ is normalization prefactor.
Therefore the corresponding $\chi$ should be proportional to the
density matrix of a pure maximally entangled state. In particular,
for $U=X$ we expect $\chi_{\mathrm{th}}=8|\Psi^{-}\rangle\langle
\Psi^{-}|$ and for $U=Y$ we have
$\chi_{\mathrm{th}}=8|\Psi^{+}\rangle\langle \Psi^{+}|$, where
$|\Psi^{+}\rangle=\frac{1}{\sqrt{2}}(|HV\rangle+|VH\rangle)$. The
experimental results shown in Fig.~2a are in very good agreement
with these theoretical predictions and the fidelity of the
reconstructed commutators, defined as normalized overlap of $\chi$
and $\chi_{\mathrm{th}}$,  reads $F_X=0.912\pm0.008$ and
$F_Y=0.873\pm 0.009$. The operator $\chi$ is normalized such that
$\mathrm{Tr}(\chi)=\mathrm{Tr}(C^\dagger C)=2|K|^2$.  However, the
normalization factor $K$ cannot be determined solely from the
tomographic data without some reference.

We have therefore performed additional calibration measurements. We
have introduced a temporal delay between the signal and program
photons so that their wave-packets did not overlap on the PBSs and
they behaved as independent entities. To ensure the calibration data
and signal data are obtained under identical circumstances, we have
carried out an independent calibration for each $U$. With this
calibration data at hand and taking into account that the success
probability of each programmable quantum gate is $\frac{1}{4}$ we
can normalize the data and fix $\mathrm{Tr}(\chi)$ and $|K|$. As can
be seen in Table I, the experimentally determined factors $K$
coincide with theoretical prediction within the statistical error.
The fidelities of reconstructed $\chi$ are also shown in Table I.
All fidelities exceed $0.8$ which indicates good agreement between
experimental observations and theory for all five tested
commutators.

The matrix $\chi$ actually characterizes the commutator only up to a
phase factor. Indeed, two different commutators $C_1$ and
$C_2=e^{i\phi} C_1$ would be represented by the same $\chi$, because
$C_1 \rho C_1^\dagger=C_2\rho C_2^\dagger$. However, the phase
$\phi$ does play an important role in the commutation relations.
Although we can't directly measure the overall phase, we can verify
the relative phase relations between two commutators. Consider the
commutators $C_{Z,X}=2iY$ and $C_{Z,Y}=-2iX$. From the measurements
reported so far we can infer that, with high fidelity,
$C_{Z,X}=e^{i\phi_X}2Y$ and $C_{Z,Y}=e^{i\phi_Y}2X$ where the phases
$\phi_X$, $\phi_Y$ remain undetermined. We next choose
$U=\frac{1}{\sqrt{2}}(X+Y)$. By linearity, but without making any
further assumptions, we have
$C_{Z,U}=[Z,U]=\sqrt{2}(e^{i\phi_X}Y+e^{i\phi_Y}X)$, hence $C_{Z,U}$
depends on $\phi_X-\phi_Y$. We have experimentally determined this
commutator and the result is shown in the right column of Fig. 2a.
Notice the nonzero imaginary part of the matrix $\chi$. The
reconstructed $\chi$ exhibits a high overlap (fidelity
$0.913\pm0.009$) with the maximally entangled state
$|\Psi_U\rangle=\frac{1}{\sqrt{2}}(|HV\rangle-i|VH\rangle)$. One can
easily check that $|\Psi_U\rangle=\frac{1}{\sqrt{2}}(Y-X)\otimes I
|\Phi^{+}\rangle$. The relative phase factor $-1$ between Pauli
operators $X$ and $Y$ is consistent with the theoretically expected
relationship $e^{i(\phi_X-\phi_Y)}=-1$. The measurement thus
corroborates the expected phase relationship between the commutators
$C_{Z,X}$ and $C_{Z,Y}$.

\begin{table}[!b!]
\caption{
The fidelities $F$ and normalization factors $K$ for the five tested commutation relations $[Z,U]$ are listed.
The normalization factors $K_{\mathrm{calib}}$ were determined from calibration measurements,
$K_{\mathrm{th}}$ represent the theoretical predictions.  }

\begin{ruledtabular}

\begin{tabular}{cccc}
$U$ & F & $K_{\mathrm{calib}}$ & $K_{\mathrm{th}}$ \\
\hline
$X$ &   $0.912\pm 0.008$    & $1.98\pm 0.03$     &  2.00 \\
$\frac{Y+Z}{\sqrt{2}}$  &   $0.800 \pm 0.019$   & $1.40 \pm 0.04$    &  1.41 \\
$Y$ &   $0.873\pm 0.009$    & $1.98 \pm 0.03$    &  2.00 \\
$\frac{X+Y}{\sqrt{2}}$ &    $0.913\pm 0.009$    & $1.95 \pm 0.03$    &  2.00 \\
$H$ &   $0.852\pm 0.016$    & $1.39 \pm 0.04$    &  1.41

\end{tabular}
\end{ruledtabular}
\end{table}

\begin{table}[!b!]
\caption{The same as Table I but results for anti-commutation
operations $\{Z,U\}$ are shown.}

\begin{ruledtabular}

\begin{tabular}{cccc}
$U$ & F & $K_{\mathrm{calib}}$  & $K_{\mathrm{th}}$ \\
\hline
I & $0.910\pm 0.008$    & $1.97 \pm 0.03 $      &  2.00 \\
Z & $0.897\pm 0.009$    & $1.95 \pm 0.03$    &  2.00 \\
$\frac{Y+Z}{\sqrt{2}}$ &    $0.901\pm 0.011$    & $1.36 \pm 0.03$    &  1.41\\
H & $0.869\pm 0.011$    & $1.39 \pm 0.03$    & 1.41
\end{tabular}
\end{ruledtabular}
\end{table}

 After testing the commutation relations,
we have proceeded to verify the anti-commutation properties of Pauli
operators.
 For this purpose, we have changed the two-photon program state to $|\Phi^{-}\rangle$
and performed complete tomographic characterization of the
anti-commutator $A_{Z,U}=\{Z,U\}$ for four different operators $I$,
$Z$, $H$, and $\frac{1}{\sqrt{2}}(Y+Z)$. Examples of results of
tomographic reconstruction are given in Fig.~2b. Since $A_{Z,I}=2Z$
and $A_{Z,Z}=2I$, the corresponding $\chi$ are proportional to
density matrices of  Bell states $|\Phi^{-}\rangle$ and
$|\Phi^{+}\rangle$, respectively. The experimental results shown in
Fig. 2b are in good agreement with theory, as witnessed by high
fidelities of the reconstructed completely positive maps,
$F_{A_{Z,I}}=0.910\pm 0.008$ and $F_{A_{Z,Z}}=0.897\pm 0.009$. The
reconstructed $\chi$ representing $A_{Z,H}$ and $A_{Z,Z}$ are very
similar, which is not surprising because $\{Z,H\}=\sqrt{2}I$. The
difference is only in the normalization of $\chi$ that was
determined from calibration measurements and reflects the different
amplitudes of $A_{Z,Z}$ and $A_{Z,H}$. All four experimentally
evaluated normalization factors $|K|$ and fidelities are summarized
in Table II.

 \begin{figure}[!t!]
 \includegraphics[width=0.95\linewidth]{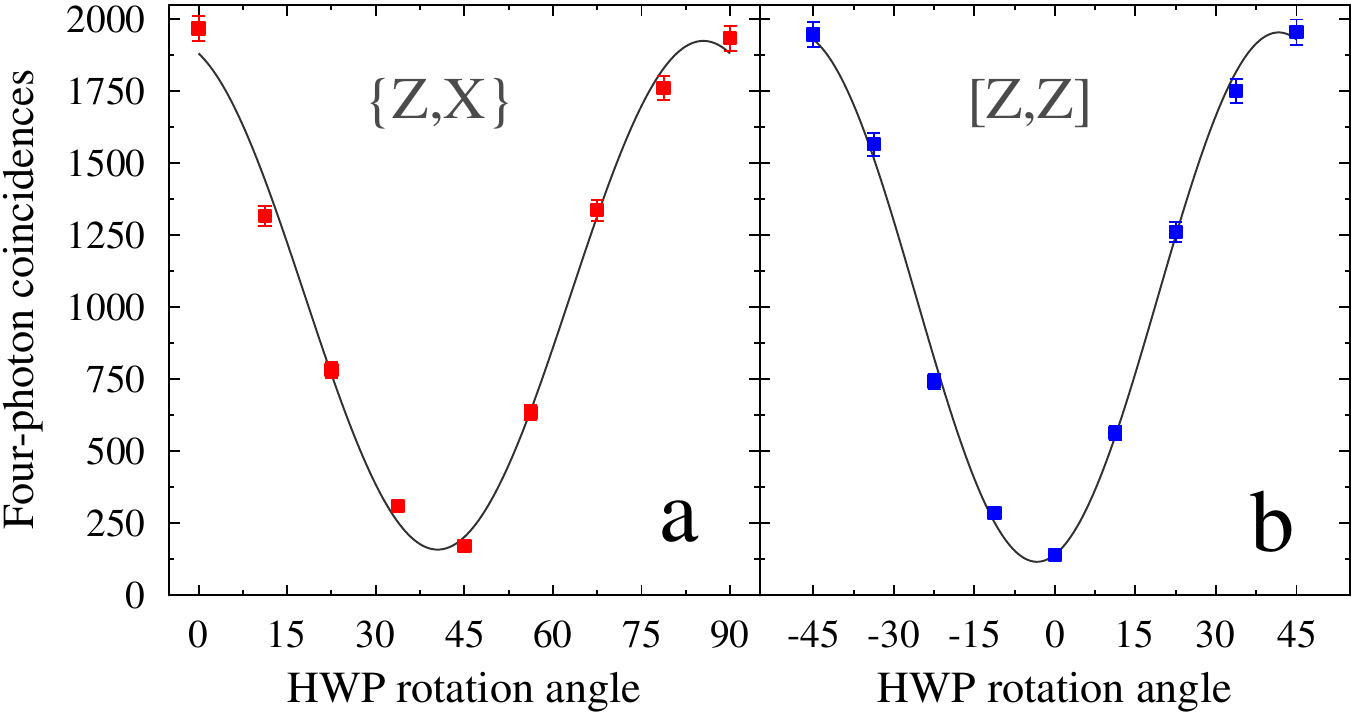}
 \caption{Four-photon coincidence dips. We plot dependence of the four-photon coincidence counts on transformation $U$ parameterized by the rotation angle of HWP. The dips directly demonstrate the  anti-commutativity of $Z$ and $X$ (a) and commutativity of $Z$ operator with itself (b). The coincidence counts were measured for six input states $|H\rangle$,
 $|V\rangle$, $|D\rangle$, $|A\rangle$, $|R\rangle$, $|L\rangle$ and summed up. The squares are experimental data and the solid line represents best sinusoidal fit. 
 The error bars are determined from Poissonian statistics.  }
 \end{figure}

 Finally, we directly observed the anti-commutativity of two different Pauli operators.
Since e.g. $\{Z,X\}=0$, we should not observe any four-photon
coincidences when $U=X$ and program photons are prepared in state
$|\Phi^{-}\rangle$. By rotating the central half-wave plate by angle
$\alpha$ we  set $U=\cos(2\alpha) Z+ \sin(2\alpha)X$ and measured
the dependence of total number of coincidences on $\alpha$, that in
theory should be proportional to $\cos^2(2\alpha)$. The results are
plotted in Fig. 3a together with a fit by a sinusoidal function. We
can clearly see the dip in coincidences at $\alpha \approx 45^\circ$
which is a direct manifestation of the anti-commutativity of $Z$ and
$X$. The visibility of the dip obtained from the fit reads
$\mathcal{V}=84.6\pm 0.5\%$. 
We have also tested the relation $\{Z,Y\}=0$. The results were very
similar and are not presented here. Finally, we have directly
checked that $Z$ commutes with itself, $[Z,Z]=0$. We prepared
program photons in state $|\Psi^{-}\rangle$ and scanned $\alpha$
over the interval $[-45^\circ,45^\circ]$. The result can be seen in
Fig. 3b with the visibility $\mathcal{V}=88.7\pm 0.5\%$.

  The coincidence dips plotted in Fig.~3 are slightly
shifted with respect to the theoretically expected positions. We
attribute this effect to the imperfections of the optical elements
employed in the experiment. Another source of experimental errors is noise in the state
generation and the imperfect overlap of photons on the PBSs, which
reduces the visibility of four-photon interference. 

  In summary, we have devised and implemented a linear optical scheme that enables
direct observation and testing of quantum commutation relations for
Pauli operators. In this way we can directly probe the
non-commutativity of quantum operators corresponding to different
physical quantities, which is one of the cornerstones of quantum
physics. The demonstrated scheme also represents an advanced
programmable quantum gate, where the type of operation (commutation
or anti-commutation) is decided by the state of two-photon program
register. By altering the program state, a whole class of operations
can be realized, including all linear combinations of $[Z,U]$ and
$\{Z,U\}$. Thus, besides being of fundamental interest, our work may
also find applications in quantum information processing.

\acknowledgments We acknowledge the financial support from the CAS,
the National Fundamental Research Program of China under Grant
No.2006CB921900, and the NNSFC. J.F. acknowledges support  by Czech
Ministry of Education under projects LC06007 and MSM6198959213 and by GACR under project
GA202/09/0747.


\begin{thebibliography}{99}

\bibitem{Peres95}
A. Peres, \emph{Quantum Theory: Concepts and Methods}, (Kluwer Academic Publishers, Dordrecht, 1995).



 \bibitem{Zavatta04}
A. Zavatta, S. Viciani, and M. Bellini, Science \textbf{306}, 660
(2004).



\bibitem{Ourjoumtsev06}
A. Ourjoumtsev, R. Tualle-Brouri, J. Laurat, and Ph. Grangier,
Science \textbf{312}, 83 (2006).


\bibitem{Neergaard-Nielsen06}
J.S. Neergaard-Nielsen {\it et al.}, Phys. Rev. Lett. \textbf{97},
083604 (2006).


\bibitem{Wakui07}
K. Wakui, H. Takahashi, A. Furusawa, and M. Sasaki, Opt. Express
\textbf{15}, 3568 (2007).

\bibitem{Parigi07} V. Parigi, A. Zavatta, M.S. Kim, and M. Bellini,
Science \textbf{317}, 1890 (2007).

\bibitem{Kim08} M. S. Kim {\it et al.},
Phys. Rev. Lett. \textbf{101}, 260401 (2008).

\bibitem{Zavatta09} A. Zavatta {\it et al.},
Phys. Rev. Lett. \textbf{103}, 140406 (2009).


\bibitem{Nielsen00}
M.A. Nielsen and I.L. Chuang, \emph{Quantum Computation and Quantum
Information,} (Cambridge University Press, Cambridge, 2000).

 \bibitem{Nielsen97} M.A. Nielsen and I. L. Chuang, Phys. Rev. Lett. \textbf{79}, 321 (1997).

\bibitem{Vidal02} G. Vidal, L. Masanes, and J. I. Cirac, Phys. Rev. Lett. \textbf{88}, 047905 (2002).

\bibitem{Micuda08} M. Mi\v{c}uda, M. Je\v{z}ek, M. Du\v{s}ek, and J. Fiur\'{a}\v{s}ek,
 Phys. Rev. A \textbf{78}, 062311 (2008).

\bibitem{Kwiat95}
P.G. Kwiat {\it et al.},
Phys. Rev. Lett. \textbf{75}, 4337-4341 (1995).

\bibitem{Pan08}
Jian-Wei Pan {\it et al.}, arXiv:0805.2853v1.

\bibitem{Poyatos97} J. F. Poyatos, J. I. Cirac and P. Zoller,
Phys. Rev. Lett. \textbf{78}, 390 - 393 (1997).

\bibitem{OBrien04}
J. L. O'Brien {\it et al.},
Phys. Rev. Lett. \textbf{93}, 080502 (2004).


\bibitem{Lobino08} M. Lobino {\it et al.},
Science \textbf{322}, 5901, 563-566 (2008).


\bibitem{Reck94} M. Reck, A. Zeilinger, H.J. Bernstein and P. Bertani,
Phys. Rev. Lett. \textbf{73}, 58-61 (1994).


\bibitem{Pittman01} T. B. Pittman, B. C. Jacobs, and J. D. Franson, 
Phys. Rev. A \textbf{64}, 062311 (2001).

\bibitem{Pittman02} T. B. Pittman, B. C. Jacobs, and J. D. Franson,
Phys. Rev. Lett. \textbf{88}, 257902 (2002).

\bibitem{Jamiolkowski72}
A. Jamiolkowski, Rep. Math. Phys. \textbf{3}, 275 (1972).

\bibitem{Choi75}
M. Choi, Linear Algebr. Appl. \textbf{10}, 285-290, (1975).

\bibitem{Jezek03}
M. Je\v{z}ek, J. Fiur\'{a}\v{s}ek, and Z. Hradil,
Phys. Rev. A \textbf{68}, 012305 (2003).

 \end{thebibliography}
\end{document}